\documentclass[11pt, prd, aps, amssymb, amsmath, tightenlines, twocolumn]{revtex4}

\usepackage[utf8]{inputenc}
\usepackage{graphicx}
\usepackage{xcolor}
\usepackage{amsmath}
\usepackage{amsfonts}
\usepackage{amssymb}
\usepackage{adjustbox}
\usepackage{multirow}
\usepackage{lineno}
\usepackage{tikz}
\usepackage{textcomp}
\usepackage{tabularx}
\usepackage{rotating}
\usepackage{fancyhdr}

\begin{document}

\pagestyle{plain}

\title{Emergence of kaonium as a sharp resonance in photon--photon to meson--meson cross-sections}

\author{Alireza Beygi$^1$}\email{alirezabeygi389@gmail.com}
\author{S. P. Klevansky$^2$}\email{spk@physik.uni-heidelberg.de}
\author{R. H. Lemmer$^3$}\email{deceased}

\affiliation{$^1$Independent Researcher, Ederstr. 14, 60486 Frankfurt am Main, Germany
\\
$^2$Institute for Theoretical Physics, Heidelberg University, 69120 Heidelberg, Germany
\\
$^3$Physics Department, University of the Witwatersrand, Johannesburg, 2050, South Africa
}

\date{\today}

\begin{abstract}

We calculate the binding energies of the hypothetical mesonic atom, $K^+ K^-$ (kaonium), using the $K^+ K^- \to K^+ K^-$ elastic scattering amplitude. Our findings are in line with previously reported results, which involve solving an eigenvalue equation of the Kudryavtsev--Popov type. Using chiral perturbation theory, we show that kaonium manifests itself as a sharp resonance around 992 MeV accompanying $f_0 (980)$ or $a_0 (980)$ in cross-sections for processes $\gamma \gamma \to \pi^0 \pi^0$ or $\gamma \gamma \to \pi^0 \eta$. The latter process is particularly striking: the peak at the kaonium resonance energy is highly pronounced, with the ratio of the cross-sections $\sigma (\gamma \gamma \to \pi^0 \eta) / \sigma (\gamma \gamma \to \pi^0 \pi^0) \approx 9$. Due to the short lifetime of kaonium ($\sim 10^{-18}$ s) and its small decay width ($\sim 0.4$ keV), direct detection of this exotic atom poses a significant challenge and requires high experimental resolution. However, we show that once the formation of kaonium is considered in the cross-section, a better fit to the available experimental data is obtained.

\end{abstract}

\maketitle

\section{Introduction}
\label{in}

The existence of the hadronic atom kaonium, $K^+ K^-$, for which there is currently no supporting experimental evidence, is still debated, and it remains a hypothetical exotic atom. On the theoretical side, although kaonium has not attracted as much attention as, for example, pionium ($\pi^+ \pi^-$) in the literature, there have still been attempts to address the properties of this elusive but fascinating atom. One of the first studies discussing the production of kaonium is given in \cite{OD1985}, where the author has calculated the energy shift of the ground state (1s) of kaonium due to strong interactions, using methods developed in the context of antiprotonic atoms \cite{OD1984}. Note that, assuming only Coulomb interactions to bind $K^+ K^-$, the unshifted energy of the 1s state lies at $-\alpha^2 \mu / 2 \approx -6.576$ keV, where $\alpha \approx 1 / 137$ is the fine structure constant and $\mu = m_{K^\pm} / 2$ is the reduced mass of the system with $m_{K^\pm} \approx 493.677$ MeV. The importance of considering the two scalar resonances $f_0 (980)$ and $a_0 (980)$ in studying strong interactions in the kaonium system, since their masses are close to $2 m_{K^\pm}$, is discussed in \cite{SW1993}. By interpreting $f_0 (980)$ and $a_0 (980)$ as quasibound states of $K \overline{K}$ \cite{JW1983, JW1990} and using the $K \overline{K}$ scattering length obtained via a separable potential formalism \cite{FC1988, FC1992}, the authors of \cite{SW1993} have made a phenomenological estimate of the lifetime of kaonium in the ground state, which is found to be $\sim 10^{-18}$ s. The mixing of $f_0 (980)$ and $a_0 (980)$ due to a coupling with kaonium is proposed in \cite{BK1995, SB1996}, where the authors show that Coulomb interactions are essential for the formation of kaonium. Following previously proposed assumptions, i.e., mixing kaonium with scalar mesons and considering a $K \overline{K}$ structure for scalar resonances, a rigorous calculation of the lifetime of kaonium is presented in \cite{SK2004}. To study the dynamics of the $K \overline{K}$ system, the authors of \cite{SK2004} have assumed a nonrelativistic $SU(3)_V \times SU(3)_A$ invariant effective Lagrangian, where strong interactions are generated by the exchange of vector mesons. The authors demonstrate that the resulting local, one-meson exchange potentials are sufficient to determine the scattering and bound-state properties of the $K \overline{K}$ system. The energy levels and decay widths of kaonium are determined by constructing the associated Jost function. The zeros of the Jost function, leading to an eigenvalue equation of the Kudryavtsev--Popov type \cite{AEK1979, AEK19792}, give the bound states of kaonium. Within this approach, the strong-interaction effects are incorporated through the complex scattering length for $K \overline{K}$ scattering and annihilation. The authors conclude that strong interactions shift the binding energy of Coulombic kaonium only by $\sim 3\%$; however, the lifetime of the kaonium ground state is significantly reduced, from $\sim 10^{-16}$ s for Coulombic kaonium to $\sim 10^{-18}$ s in the presence of strong interactions. Momentum-dependent interactions have been neglected in the potentials obtained in \cite{SK2004}; inclusion of these terms is considered in \cite{YZ2006} and \cite{RHL2009}, leading to a nonlocal potential in coordinate space. However, according to \cite{RHL2009}, the properties of kaonium, i.e., the energy shifts, decay widths, and its lifetime, within this approach are in line with previous estimates using the local potential approximation in \cite{SK2004}. Rather than generating meson--meson interaction vertices by the exchange of vector mesons, the authors of \cite{SPK2011} have employed a different approach developed in \cite{JAO1997}, where meson--meson potentials are extracted from the Lagrangian given by leading-order chiral perturbation theory (ChPT). Using these interactions as input for the Lippmann--Schwinger equation, one can obtain the $T$-matrix elements for a non-perturbative description of meson--meson scattering and reactions. Within this framework, the effects of strong interactions on the energy shifts and decay widths of kaonium are recalculated in \cite{SPK2011}. The reported results are in line with those in \cite{SK2004}. More recently, an indication of the 2p state of kaonium is discussed in \cite{PL2020}, based on the analysis of the data of the CMD-3 experiment on the $e^+ e^- \to K^+ K^-$ process. The exclusive production of kaonium via photon fusion in ultraperipheral hadron collisions, with estimates of expected yields at the LHC, has been studied in \cite{DdE2025}.

In the current paper, rather than focusing mainly on the energy shifts, decay widths, or the lifetime of kaonium, we investigate how this elusive, exotic atom manifests its presence in the cross-sections for processes $\gamma \gamma \to \pi^0 \pi^0$ or $\gamma \gamma \to \pi^0 \eta$. To this end, we employ a version of ChPT, proposed in \cite{JAO1997, JAO1998}, to calculate the cross-sections including the formation of kaonium. For this, a solid knowledge of the strong-interaction scattering amplitudes is required. We have already presented detailed calculations of the respective amplitudes for these processes, in the context of \cite{JAO1997, JAO1998}, in a previously published paper \cite{SPK2025}. These amplitudes need modifications to incorporate the formation of kaonium, which is the subject of our current paper. We show that in cross-sections kaonium emerges as a sharp peak around 992 MeV, accompanying $f_0 (980)$ or $a_0 (980)$ resonances. The cross-section curve, once the kaonium formation is included, provides a better fit to the available experimental data.

This paper is organized as follows. In Sec.~\ref{be}, we calculate the binding energy of kaonium using the $K^+ K^-$ elastic scattering amplitude. This involves the inclusion of isospin breaking due to the mass difference $K^0 - K^\pm$ and attractive Coulomb interactions, which we present in Sec.~\ref{2A} and Sec.~\ref{2B}, respectively. The kaonium binding energies obtained in Sec.~\ref{2B} are in line with those reported in \cite{SK2004, SPK2011}. Calculations of cross-sections are given in Sec.~\ref{cs}. The process $\gamma \gamma \to \pi^0 \pi^0$ is discussed in Sec.~\ref{3A} and $\gamma \gamma \to \pi^0 \eta$ in Sec.~\ref{3B}. For the latter, we also compare our theoretical cross-section curve with the experimental data of the JADE and Belle Collaborations \cite{JADE1990, Belle2009}. The ratio of the cross-sections for these two processes at the kaonium resonance energy is given in Sec.~\ref{3C}. We conclude and summarize in Sec.~\ref{co}.

\section{Binding energy of kaonium}
\label{be}

In the analysis of kaonium, it is essential to consider isospin breaking that stems from two different sources. There is an isospin breaking because of the meson mass difference, which induces the decay of this exotic atom. Another source of isospin breaking is due to Coulomb interactions, which lead to the binding of $K^+ K^-$. In the following, we explain how to treat these two cases.

\subsection{Inclusion of isospin breaking due to the mass difference $K^0 - K^\pm$}
\label{2A}

Isospin breaking due to the kaon mass difference, $\Delta = m_{K^0} - m_{K^\pm}$, within the context of $K^-$--proton interactions, has been discussed in \cite{RHD1959}. Here, we recast the formalism of \cite{RHD1959} within the context of our paper and set our notation.

First, recall how the $K^+ K^-$ physical scattering length due to strong interactions can be determined if one ignores isospin breaking. For this, we need to evaluate the amplitude of $K^+ K^-$ scattering, that is, $T_{K^+ K^-} = \langle K^+ K^- | T | K^+ K^- \rangle$. Note that for
\begin{equation}
K = \left( \begin{matrix}
K^+ \\
K^0
\end{matrix} \right), \quad \overline{K} = \left( \begin{matrix}
\overline{K}^0 \\
-K^-
\end{matrix} \right),
\end{equation}
the isoscalar and isovector combinations giving $| I = 0, I_3 = 0 \rangle$ and $| I = 1, I_3 = 0 \rangle$ read as
\begin{equation}
\begin{aligned}
\left | \left ( K \overline{K} \right )^0 \right \rangle &= -\frac{1}{\sqrt 2} \biggl ( \left | K^+ K^- \right \rangle + \left | K^0 \overline{K}^0 \right \rangle \biggr ), \\
\left | \left ( K \overline{K} \right )^1 \right \rangle &= -\frac{1}{\sqrt 2} \biggl ( \left | K^+ K^- \right \rangle - \left | K^0 \overline{K}^0 \right \rangle \biggr ),
\label{iic}
\end{aligned}
\end{equation}
and the corresponding inverse relations are
\begin{equation}
\begin{aligned}
\left | K^+ K^- \right \rangle &= -\frac{1}{\sqrt 2} \biggl ( \left | \left ( K \overline{K} \right)^0 \right \rangle + \left | \left ( K \overline{K} \right)^1 \right \rangle \biggr ), \\
\left | K^0 \overline{K}^0 \right \rangle &= -\frac{1}{\sqrt 2} \biggl ( \left | \left ( K \overline{K} \right)^0 \right \rangle - \left | \left ( K \overline{K} \right)^1 \right \rangle \biggr ).
\label{iic2}
\end{aligned}
\end{equation}
Then, the scattering amplitude can be evaluated as
\begin{align}
T_{K^+ K^-} &= \left \langle K^+ K^- \left | T \right | K^+ K^- \right \rangle \nonumber \\
&= \frac{1}{2} \biggl ( \left \langle \left ( K \overline{K} \right )^0 \left | T \right | \left ( K \overline{K} \right )^0 \right \rangle \nonumber \\
&\; \; \quad \quad + \left \langle \left ( K \overline{K} \right )^1 \left | T \right | \left ( K \overline{K} \right )^1 \right \rangle \biggr ) \nonumber \\
&= \frac{1}{2} \left ( T^0_{1 1} + T^1_{1 1} \right ).
\label{tkk}
\end{align}
In the notation for $T^I_{i j}$, the indices $(i, j)$ identify the specific meson pair involved; here, 1 indicates $K \overline{K}$ in both isospin states $I = 0$ and $I = 1$; see also \cite{SPK2025}. Now, having $T_{K^+ K^-}$ at our disposal, the $K^+ K^-$ scattering length is determined by
\begin{align}
a_{K^+ K^-} &= -\frac{\left \langle K^+ K^- \left | T \right | K^+ K^- \right \rangle}{16 \pi m_K} \nonumber \\
&= \frac{1}{2} \left ( a^0_{K \overline{K}} + a^1_{K \overline{K}} \right ),
\label{psl}
\end{align}
where we have used the standard formula $a^I_{K \overline{K}} = - T^I_{1 1} / 8 \pi \sqrt{s}$ as $s \to 4 m_K^2$ \cite{JLP1971}.

As is pointed out in \cite{SPK2011}, the expression obtained in (\ref{psl}) is only valid if we ignore the isospin breaking due to the meson mass difference $\Delta = m_{K^0} - m_{K^\pm}$; if not, the $K^+ K^- \to K^0 \overline{K}^0$ charge transfer channel is closed \cite{RHD1959} and the $K^+ K^-$ scattering length in (\ref{psl}) needs to be modified. To perform this modification, first note that outside the strong-interaction zone, $r \sim d$, the behavior of the charged $K^+ K^-$ and neutral $K^0 \overline{K}^0$ radial wave functions, $v_p$ and $v_n$, can be determined by solving the corresponding Schr\"odinger equations, leading to: $v_p = \sin (k r + \delta_p) / \sin \delta_p$ and $v_n = C_n e^{i k_n r}$, where $\delta_p$ is the (possibly complex) phase shift; see also \cite{HAB1949} for a detailed and analogous discussion within the context of nuclear scattering. The wave numbers $k$ and $k_n$ are related by total energy conservation: $k^2 / 2 \mu + 2 m_{K^\pm} = k_n^2 / 2 \mu + 2 m_{K^0}$, where $\mu$ is the average reduced mass and $\epsilon = k^2 / 2 \mu$ is the center-of-mass energy. Thus, $k_n = i \sqrt{2 \mu (2 \Delta - \epsilon)} = i k_0$. The physical charged and neutral channel wave functions $v_p$ and $v_n$ can be written in terms of the solutions of good isospin $u_0$ and $u_1$ outside the strong-interaction zone as $u_0 = -(v_p + v_n) / \sqrt{2}$ and $u_1 = -(v_p - v_n) / \sqrt{2}$. Using $u_0$ and $u_1$, the joining relations at $r = d$ outside the strong-interaction zone are
\begin{equation}
\begin{aligned}
\xi_0 u_0 &= -\frac{1}{\sqrt{2}} \left( \xi_p v_p + \xi_n v_n \right), \\
\xi_1 u_1 &= -\frac{1}{\sqrt{2}} \left( \xi_p v_p - \xi_n v_n \right),
\label{llo}
\end{aligned}
\end{equation}
where $\xi$'s are logarithmic derivatives, i.e., $\xi_p = v_p^{'} / v_p$, etc. The equations in (\ref{llo}) can then be rewritten as \cite{SK2004},
\begin{equation}
\left( \begin{matrix}
\xi_p - \xi_0 & \xi_n - \xi_0 \\
\xi_p - \xi_1 & -(\xi_n - \xi_1)
\end{matrix} \right) \left( \begin{matrix}
v_p \\
v_n
\end{matrix} \right) = 0.
\label{xima}
\end{equation}
Non-trivial solutions for $v_{p, n}$ exist provided that the determinant of the coefficient matrix vanishes, which leads to: $(\xi_p + \xi_n)(1 / \xi_0 + 1 / \xi_1) / 2 = 1 + \xi_p \xi_n / \xi_0 \xi_1$. Now, recall that at the zero-range approximation, i.e., $r = d \to 0$, the logarithmic derivatives are known in terms of the scattering lengths of good isospin: $\xi_0 = - 1 / a^0_{K \overline{K}}$, $\xi_1 = - 1 / a^1_{K \overline{K}}$, $\xi_p = - 1 / a_p$, and also $\xi_n = i k_n = - k_0$, where the zero-momentum limit, that is, $k \to 0$, is assumed. Substituting these $\xi$'s into the expression of the vanishing determinant yields
\begin{equation}
a_p = \frac{a_{K^+ K^-} - k_0 a^0_{K \overline{K}} a^1_{K \overline{K}}}{1 - k_0 a_{K^+ K^-}},
\label{app}
\end{equation}
which is also quoted in \cite{RHL2009, SPK2011} without a derivation. In (\ref{app}), $k_0 = \sqrt{2 m_{K^0} \Delta}$. Thus, instead of $a_{K^+ K^-}$ in (\ref{psl}), the correct formula for the $K^+ K^-$ scattering length that includes isospin breaking due to the kaon mass difference $\Delta$ is given in (\ref{app}). Note that in the limit $\Delta \to 0$, $a_p$ coincides with $a_{K^+ K^-}$ in (\ref{psl}).

To obtain the associated low-energy amplitude, first recall that the asymptotic form of the scattering wave function along $z$ is $\psi \sim e^{i k z} + f e^{i k r} / r$, as $r \to \infty$ and $f$ is the scattering amplitude; see \cite{AM1962}. For the $s$-wave contribution at low energies, this reduces to: $\psi_s \sim \sin k r / k r + f_p e^{i k r} / r$. Thus, $v_p = r \psi_s \sim \sin k r / k + T_p e^{i k r} / 8 \pi \sqrt{s}$. The values of $\hat{T}_p$ and $\hat{T}_n$ of the charged and neutral scattering channels defined through $v_p = \sin k r / k + \hat{T}_p e^{i k r}$ and $v_n = \hat{T}_n e^{i k_n r}$ still satisfy (\ref{xima}), that is,
\begin{equation}
\left( \begin{matrix}
\xi_p - \xi_0 & \xi_n - \xi_0 \\
\xi_p - \xi_1 & -(\xi_n - \xi_1)
\end{matrix} \right) \left( \begin{matrix}
\hat{T}_p \\
\hat{T}_n
\end{matrix} \right) = 0,
\end{equation}
with the revised value of $\xi_p = (1 + i k \hat{T}_p) / \hat{T}_p$. Thus, $\xi_p = (1 + i k \hat{T}_p) / \hat{T}_p = - 1 / a_p$, then $\hat{T}_p = - a_p / (1 + i k a_p)$. Using (\ref{psl}) and (\ref{app}), $T_p$ can be written as
\begin{align}
T_p = &-8 \pi \sqrt{s} \nonumber \\
&\times \frac{(a_0 + a_1)/2 + i k_n a_0 a_1}{1 + i (k + k_n)(a_0 + a_1)/2 - k k_n a_0 a_1},
\label{Tp}
\end{align}
where we have simplified the notation by writing $a^I_{K \overline{K}} = a_I$. The relation in (\ref{Tp}) is in agreement with the result reported in \cite{RHD1959} in the context of the scattering of $K^-$ by a proton. A respective relation can be obtained for $\hat{T}_n$, by noting that $\hat{T}_n / \hat{T}_p = - (\xi_p - \xi_0) / (\xi_n - \xi_0)$.

In the following, we summarize the formulas for the associated low-energy phase shift, and the $S$-, and $T$-matrix elements:
\begin{widetext}
\begin{equation}
\begin{aligned}
k \cot \delta_p &= -\frac{1}{a_p} = -\frac{1 - k_0 a_{K^+ K^-}}{a_{K^+ K^-} - k_0 a_0 a_1}, \\
S_p &= e^{2 i \delta_p} = \frac{k \cot \delta_p + i k}{k \cot \delta_p - i k} = \frac{1 - i k a_p}{1 + i k a_p} = \frac{1 - k_0 a_{K^+ K^-} - i k \left( a_{K^+ K^-} - k_0 a_0 a_1 \right)}{1 - k_0 a_{K^+ K^-} + i k \left( a_{K^+ K^-} - k_0 a_0 a_1 \right)} \\
&= 1 + 2 i k \frac{T_p}{8 \pi \sqrt{s}}, \\
\frac{T_p}{8 \pi \sqrt{s}} &= \frac{1}{k \cot \delta_p - i k} = -\frac{a_p}{1 + i k a_p}, \\
T_n &= \frac{a_0 - a_1}{2 \left( a_{K^+ K^-} - k_0 a_0 a_1 \right)} T_p.
\label{pst1}
\end{aligned}
\end{equation}
\end{widetext}

\subsection{Inclusion of strong plus Coulomb interactions}
\label{2B}

The approach to combine the effects of strong and Coulomb interactions follows the standard procedure of joining the respective two wave functions, one corresponding to the region where strong interactions dominate and the other to the region where Coulomb interactions dominate, at an intermediate region. This approach has been discussed in \cite{LL1944, HAB1949, JDJ1950} for proton--proton scattering, and was later applied to $K^-$--proton scattering in \cite{RHD1960}, and to kaonium formation in \cite{SK2004}. In the following, we reformulate this method in the context of our paper to determine the correct form of the amplitude for $K^+ K^-$ scattering, taking into account both strong and Coulomb interactions.

First, recall that in Sec.~\ref{2A}, we introduced: $v_p = \cos k r + \sin k r \cot \delta_p$. Now, the wave function $v$, where $v \neq v_p$, outside the strong-interaction zone is modified to read \cite{LL1944, HAB1949, JDJ1950},
\begin{equation}
v = C_0 \Bigl( G + F \cot \delta \Bigr),
\label{modv}
\end{equation}
where $\delta \neq \delta_p$ is to be interpreted as the revised strong phase shift in the presence of the Coulomb field. In (\ref{modv}), $C_0$ is the normalization constant for Coulomb wave functions, that is, $C_0 = \sqrt{2 \pi \eta / (e^{2 \pi \eta} - 1)}$, where the Coulomb parameter $\eta = \mu \alpha / k$, and the regular and irregular (logarithmic) Coulomb wave functions $F$ and $G$ for $s$-waves are defined in terms of the incoming wave in the Coulomb field, $f^{(-)}_c$, as \cite{FLY1936, RGN1966},
\begin{equation}
f^{(-)}_c = e^{i \sigma} \left( G - i F \right),
\label{fci}
\end{equation}
where $f^{(-)}_c$ is proportional to the Whittaker function, i.e., $f^{(-)}_c = e^{\pi \eta / 2} W_{i \eta, 1 / 2}(2 i k r)$, and $\sigma = \text{arg} \; \Gamma(1 + i \eta)$ is the Coulomb phase shift. Note that if Coulomb interactions are turned off at large distances, $v$ in (\ref{modv}) coincides with $v_p$. To see this, first note that $f^{(-)}_c \sim e^{-i (k r - \eta \ln 2 k r)}$. Thus, based on (\ref{fci}), at large distances $F$ and $G$ behave as \cite{FLY1936},
\begin{equation}
\begin{aligned}
F &\sim \sin \left( k r - \eta \ln 2 k r + \sigma \right), \\
G &\sim \cos \left( k r - \eta \ln 2 k r + \sigma \right).
\end{aligned}
\label{FGe}
\end{equation}
Removing Coulomb interactions implies $\eta = 0$, leading to: $v \to v_p$.

Now, for continuity of the scattering wave function, we match the logarithmic derivatives of $v_p$ and $v$ at some joining radius $d$ outside the strong-interaction zone:
\begin{equation}
\frac{v^{'}_p}{v_p} \Biggr|_{r = d} = \frac{v^{'}}{v} \Biggr|_{r = d}.
\end{equation}
Note that the range of strong interactions is $1 / m_K \sim 0.4$ fm, however, the length scale over which Coulomb interactions operate is the Bohr radius of $1 / \mu \alpha \sim 110$ fm, thus the joining radius $d$ satisfies: $1 / m_K \lesssim d \ll 1 / \mu \alpha$. This implies that the joining radius $d$ is such that while the strong-interaction wave function has already reached its asymptotic form $v_p$ at \textit{large} $r$, the Coulomb waves have hardly had space to crawl away from the origin; thus, we can use a small argument approximation to simplify their corresponding wave functions. For this, note that at small distances $F$ and $G$ have the form \cite{LL1944},
\begin{equation}
\begin{aligned}
&F \sim C_0 k r, \\
&G \sim C_0^{-1} \biggl\{ 1 + 2 \eta k r \Bigl[ \ln 2 \eta k r + 2 \gamma - 1 + h \left( \eta \right) \Bigr] \biggr\},
\end{aligned}
\end{equation}
where $h(\eta) = \text{Re} \, \Psi(1 + i \eta) - \ln \eta$, $\Psi$ is the digamma function, and $\gamma$ is Euler's constant. Now, the logarithmic derivatives of $v_p$ and $v$ at $r = d$ read
\begin{equation}
\begin{aligned}
&\frac{v^{'}_p}{v_p} \Biggr|_{r = d} = - \frac{\alpha_p}{1 - \alpha_p d}, \\
&\frac{v^{'}}{v} \Biggr|_{r = d} = 2 \mu \alpha \Bigl[ \ln 2 \mu \alpha d + 2 \gamma + h \left( \eta \right) \Bigr] + C_0^2 k \cot \delta \\
&\times \frac{1}{1 + 2 \mu \alpha d \Bigl[ \ln 2 \mu \alpha d + 2 \gamma + h \left( \eta \right) \Bigr] + C_0^2 k d \cot \delta},
\end{aligned}
\label{mat3}
\end{equation}
where $\alpha_p = 1 / a_p$, with $a_p$ given in (\ref{app}). The matching relation leads to
\begin{equation}
- \alpha_p = 2 \mu \alpha \Bigl[ \ln 2 \mu \alpha d + 2 \gamma + h \left( \eta \right) \Bigr] + C_0^2 k \cot \delta.
\label{alphi}
\end{equation}
Changing $\alpha \to -\alpha$ for attractive Coulomb interactions, (\ref{alphi}) becomes
\begin{widetext}
\begin{equation}
k \cot \delta = \frac{1 - e^{-2 \pi \mu \alpha / k}}{2 \pi \mu \alpha / k} \biggl\{ 2 \mu \alpha \Bigl[ \ln 2 k d + 2 \gamma + \text{Re} \, \Psi \left( 1 - i \mu \alpha / k \right) \Bigr] - \alpha_p \biggr\}.
\label{cottd}
\end{equation}
\end{widetext}
Note that when the Coulomb field is removed, that is, $\alpha \to 0$, the coefficient expression in (\ref{cottd}) behaves as $k (1 - e^{-2 \pi \mu \alpha / k}) / \pi \sim 2 \mu \alpha$, thus (\ref{cottd}) gives $k \cot \delta \to -\alpha_p$, which correctly reproduces the first relation in (\ref{pst1}).

\begin{figure*}[!htbp]

\centering

\includegraphics[scale = 1.1]{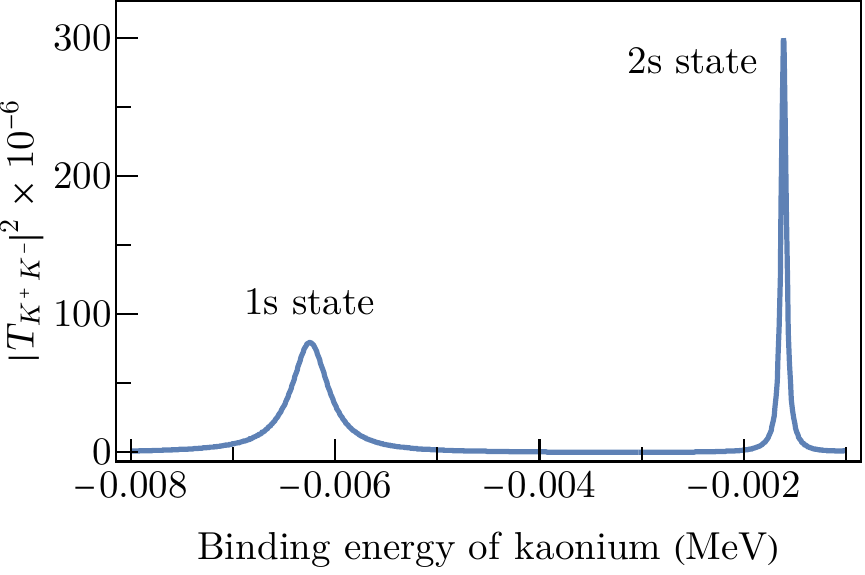}

\caption{Resonances in the $K^+ K^- \to K^+ K^-$ elastic scattering amplitude squared, $|T_{K^+ K^-}|^2 \times 10^{-6}$, where $T_{K^+ K^-}$ is given in (\ref{esT}), corresponding to the formation and decay of the 1s and 2s bound states of kaonium.}

\label{TB}
\end{figure*}

Now, the associated $S$-matrix element, which replaces $S_p$ in (\ref{pst1}), reads
\begin{equation}
S_p \to S_{k^+ K^-} = e^{2 i \left( \sigma + \delta \right)} = e^{2 i \sigma} \frac{k \cot \delta + i k}{k \cot \delta - i k},
\label{smm}
\end{equation}
with $k \cot \delta$ given in (\ref{cottd}). For the total center-of-mass energy below the $K^+ K^-$ threshold at $2 m_K$, $s = P_0^2 \leq 4 m_K^2$, $k = \sqrt{P_0^2 - 4 m_K^2} / 2$ is pure imaginary, and the correct analytic continuation is $k = + i \sqrt{4 m_K^2 - P_0^2} / 2$, which also implies $\sigma = 0$. The $K^+ K^- \to K^+ K^-$ elastic scattering amplitude, which replaces $T_p$ in (\ref{Tp}), is determined by
\begin{align}
T_p \to T_{\text{el.}} = T_{K^+ K^-} &= 8 \pi \sqrt{s} \frac{S_{K^+ K^-} - 1}{2 i k} \nonumber \\
&= \frac{8 \pi \sqrt{s}}{k \cot \delta - i k}.
\label{esT}
\end{align}
We have shown $|T_{K^+ K^-}|^2$ in Fig.~\ref{TB}, with the strong $K^+ K^-$ scattering length $a_p = (2.688 - 1.896 i) m_K^{-1}$ and $d \approx 2.2 m_K^{-1}$ taken from \cite{SPK2011}.
The resonance peaks given here by the amplitude are in line with the estimated kaonium binding energies reported in \cite{SK2004, SPK2011}, which correspond to the pole positions given by the Kudryavtsev--Popov equation.

\begin{figure*}[!htbp]

\centering

\includegraphics[scale = 1.1]{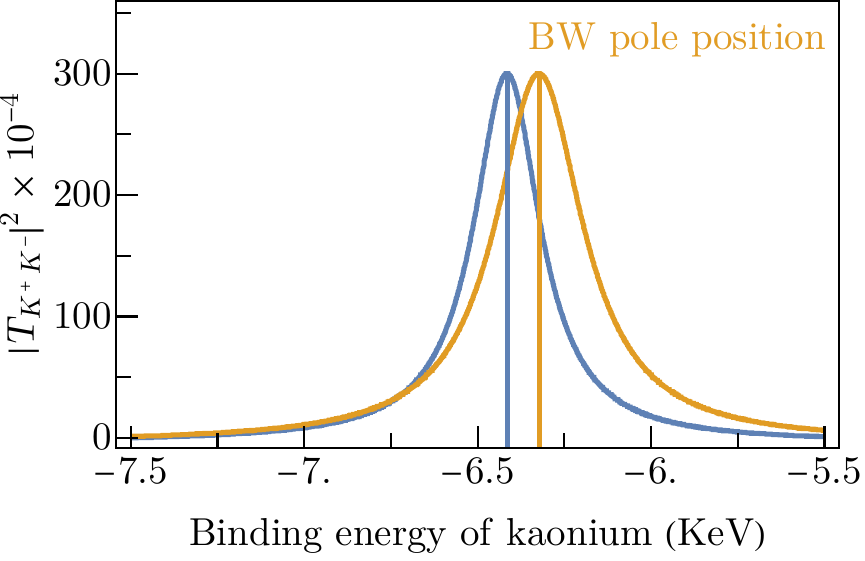}

\caption{The lowest 1s kaonium resonance in the $K^+ K^- \to K^+ K^-$ elastic scattering amplitude squared, $|T_{K^+ K^-}|^2 \times 10^{-4}$, where $T_{K^+ K^-}$ is given in (\ref{esT}), is plotted on an enlarged scale relative to that used in Fig.~\ref{TB} (blue curve). The associated Breit--Wigner (BW) curve obtained from (\ref{bw0}), using the pole position of the $T_{K^+ K^-}$ for its parameters, is slightly shifted to the right (orange curve).}

\label{TBB}
\end{figure*}

To conclude this section, we point out that the shape of the curve in Fig.~\ref{TB} is a perfect Breit--Wigner resonance. In \cite{SPK2025}, we have given the theoretical derivation of the respective amplitude for $K \overline{K} \to K \overline{K}$. The case of $K^+ K^- \to K^+ K^-$ can be worked out in the same way, resulting in:
\begin{equation}
T^{\text{BW}}_{K^+ K^-} \approx -\frac{1}{2 M_{\text{Ka.}}} \frac{g_{\text{Ka.}}^2}{P_0 - M_{\text{Ka.}} + i \Gamma_{\text{Ka.}} / 2},
\label{bw0}
\end{equation}
where $M_{\text{Ka.}}$, $\Gamma_{\text{Ka.}}$, and $g_{\text{Ka.}}$ denote the mass of kaonium, its total width, and kaonium-to-$K^+ K^-$ coupling constant. For the 1s state, the resonance peak in Fig.~\ref{TB} occurs at $-0.006252$ MeV, with a respective width of 0.000448 MeV. Thus, $M_{\text{Ka.}} = 2 m_K - 0.006252 = 991.994$ MeV and $\Gamma_{\text{Ka.}} = 0.000448$ MeV. We fix $g_{\text{Ka.}} = 62.598$ MeV to match the same peak height as in Fig.~\ref{TB}. We have shown $T^{\text{BW}}_{K^+ K^-}$ in (\ref{bw0}) for these values in Fig.~\ref{TBB}, compared to the resonance curve of Fig.~\ref{TB}. The exact curve on the left is indistinguishable from the Breit--Wigner fit, which is slightly shifted to the right.

\section{Photon--photon to meson--meson cross-sections including the formation of kaonium}
\label{cs}

Due to the very short lifetime of kaonium $\sim 10^{-18}$ s, its direct experimental detection poses significant challenges. However, an indirect way of detecting kaonium would be to look for its possible manifestation in photon--photon to meson--meson cross-sections. In our previous paper \cite{SPK2025}, we have used a version of ChPT, developed in \cite{JAO1997, JAO1998}, being an appropriate realization of low-energy QCD, to give a detailed analysis of the final-state, strong-interaction, meson--meson scattering amplitudes for the processes $\pi \pi \to \pi \pi$, $\pi \pi \to K \overline{K}$, $K \overline{K} \to K \overline{K}$, and have calculated the cross-sections for $\gamma \gamma \to \pi^0 \pi^0$ and $\gamma \gamma \to \pi^0 \eta$ within this framework. These calculations use the leading-order ChPT interaction Lagrangian density \cite{JG1984, JG1985, GE1995, JAO1997}, $\mathcal{L}_2 = 1 / (12 f^2) \text{tr} \left[ (\partial_\mu \Phi \Phi - \Phi \partial_\mu \Phi)^2 + M \Phi^4 \right]$, where $f$ is the (bare) pion decay constant, tr denotes the trace in the flavor space of the $SU(3)$ matrices constructed from
\begin{equation}
\Phi = \left( \begin{smallmatrix}
\pi^0 / \sqrt 2 + \eta / \sqrt 6 & \pi^+ & K^+ \\
\pi^- & -\pi^0 / \sqrt 2 + \eta / \sqrt 6 & K^0 \\
K^- & \overline{K}^0 & -2 \eta / \sqrt 6
\end{smallmatrix} \right),
\end{equation}
which is a matrix of the meson fields, and $M = \text{diag} (m_\pi^2, m_\pi^2, 2 m_K^2 - m_\pi^2)$ is the diagonal matrix of (bare) meson masses. Working in a set of basis states of good isospin $I$, the 4-point vertices at tree level can be evaluated as $V^I_{m_1 m_2; m_3 m_4} = V^I_{m_3 m_4; m_1 m_2} = \left \langle I, m_1 m_2 \left | \mathcal{L}_2 \right | I, m_3 m_4 \right \rangle = V^I_{i j}$. In abbreviated notation $V^I_{i j}$, the indices $(i, j) = (1, 2)$ label the meson pairs as: 1 corresponds to $K \overline{K}$ for both $I = 0$ and $I = 1$, 2 corresponds to $\pi \pi$ for $I = 0$ (or $I = 2$) and to $\pi^0 \eta$ for $I = 1$. Using $V^I_{i j}$, which are, in general, off-shell, one can obtain the associated scattering amplitudes $T^I_{i j}$ in terms of coupled Lippmann--Schwinger integral equations. However, as demonstrated in \cite{JAO1997}, when dealing with $s$-wave scattering, since the off-shell parts of $V^I_{i j}$ can be absorbed as a renormalization factor, $V^I_{i j}$ can be replaced by their on-shell values, provided that $f$ and the meson masses take on their physical values. The on-shell values of $V^I_{i j}$ are \cite{JAO1997, SPK2011, SPK2025},
\begin{equation}
\begin{aligned}
&V^0_{1 1} = \frac{3 s}{4 f^2}, \quad V^0_{2 1} = \sqrt{\frac{3}{2}} \frac{s}{2 f^2}, \quad V^0_{2 2} = \frac{2 s - m_\pi^2}{f^2}, \\
&V^1_{1 1} = \frac{s}{4 f^2}, \quad V^1_{2 1} = -\sqrt{\frac{2}{3}} \frac{9 s - m_\pi^2 - 3 m_\eta^2 - 8 m_K^2}{12 f^2}, \\
&V^1_{2 2} = \frac{m_\pi^2}{3 f^2}, \quad V^2_{2 2} = - \frac{s - 2 m_\pi^2}{f^2}.
\end{aligned}
\label{onsv}
\end{equation}
Replacing $V^I_{i j}$ by their on-shell values reduces the coupled integral equations for $T^I_{i j}$ to coupled algebraic equations that can be solved analytically, resulting in \cite{JAO1997, JAO1998, SPK2025},
\begin{equation}
\begin{aligned}
T^I_{1 1} &= \Bigl[ \bigl ( 1 - V^I_{2 2} \Pi^I_{2 2} \bigr ) V^I_{1 1} + V^I_{1 2} \Pi^I_{2 2} V^I_{2 1} \Bigr] / D^I, \\
T^I_{1 2} &= V^I_{1 2} / D^I = T^I_{2 1}, \\
T^0_{2 2} &= \left ( T^0_{1 1} \right )_{1 \leftrightarrow 2}, \\
T^2_{2 2} &= V^2_{2 2} / \bigl ( 1 - V^2_{2 2} \Pi^2_{2 2} \bigr ),
\end{aligned}
\label{e:T11}
\end{equation}
where $D^I$ is given by
\begin{align}
D^I = \bigl ( 1 - V^I_{1 1} \Pi^I_{1 1} \bigr ) \bigl ( 1 &- V^I_{2 2} \Pi^I_{2 2} \bigr ) \nonumber \\
&- V^I_{1 2} \Pi^I_{2 2} V^I_{2 1} \Pi^I_{1 1},
\label{e:roots}
\end{align}
and $\Pi^I_{i i}$, indicating the meson loop diagrams, reads
\begin{equation}
i \Pi^I_{i i} = \epsilon \int \frac{d^4 l}{( 2 \pi )^4} \frac{1}{l^2 - m^2_a} \frac{1}{( l + P_0 )^2 - m^2_b},
\label{e:Pi}
\end{equation}
with $\epsilon$ taking the value $1 / 2$ for two identical and 1 for two different mesons \cite{JLP1971}, and $m_{a, b}$ being the masses of the mesons in the loop. Notationally, we can also write $\Pi^I_{1 1} = \Pi_{K \overline{K}}$ for both $I = 0$ and $I = 1$, $\Pi^I_{2 2} = \Pi_{\pi \pi}$ for $I = 0$ (or $I = 2$), and $\Pi^I_{2 2} = \Pi_{\pi^0 \eta}$ for $I = 1$. Note that the integral in (\ref{e:Pi}) needs regularization as it diverges at large four-momenta. This requires introducing an $O(4)$ cutoff parameter $\Lambda$ to regularize this integral. In \cite{SPK2025}, we have provided a detailed analysis and evaluation of $\Pi^I_{i i} = \Pi$, where its regularized form for $s > (m_a + m_b)^2$ reads \cite{SPK2011, SPK2025},
\begin{align}
\Pi = \frac{\epsilon}{( 4 \pi )^2} \biggl [ &\frac{m^2_a}{m^2_a - m^2_b} \ln \left ( 1 + \Lambda^2 / m^2_a \right ) \nonumber \\
&- \frac{m^2_b}{m^2_a - m^2_b} \ln \left ( 1 + \Lambda^2 / m^2_b \right) - L_{a b} \biggr ],
\label{e:Piclosed}
\end{align}
where
\begin{align}
&L_{a b} = - 1 - \frac{1}{2} \left ( \frac{m^2_a + m^2_b}{m^2_a - m^2_b} - \frac{m^2_a - m^2_b}{s} \right ) \nonumber \\
&\times \ln \left ( m^2_a / m^2_b \right ) \nonumber \\
&+ \sqrt{f_{a b}} \Biggl \{ \tanh^{-1} \Biggl [ \frac{\sqrt{f_{a b}}}{1 - \left ( m^2_a - m^2_b \right ) / s} \Biggr ] \nonumber \\
&+ \tanh^{-1} \Biggl [ \frac{\sqrt{f_{a b}}}{1 + \left ( m^2_a - m^2_b \right ) / s} \Biggr ] \Biggr \} - i \pi \sqrt{f_{a b}},
\label{labab}
\end{align}
and
\begin{align}
\sqrt{f_{a b}} = &\left ( 1 - \left( m_a - m_b \right )^2 / s \right )^{1 / 2} \nonumber \\
&\times \left ( 1 - \left ( m_a + m_b \right )^2 / s \right )^{1 / 2} = 2 p_{a b} / \sqrt s,
\label{e:Lab}
\end{align}
with $p_{a b}$ denoting the magnitude of the meson's three-momentum. A closed form for $\Pi_{\pi^0 \eta}$ can be obtained by setting $m_a = m_\pi$ and $m_b = m_\eta$ in (\ref{e:Piclosed}). When the masses are equal, i.e., $m_a = m_b = m$, $\Pi$ becomes \cite{SPK2011, SPK2025},
\begin{align}
\Pi = \frac{\epsilon}{( 4 \pi )^2} \Bigl [ 1 &+ \ln \left ( 1 + \Lambda^2 / m^2 \right ) \nonumber \\
&+ m^2 / \left ( m^2 + \Lambda^2 \right ) - 2 J \Bigr ],
\label{RPi}
\end{align}
where
\begin{align}
&J = \sqrt{- f} \cot^{-1} \sqrt{- f} \; \theta \left ( 4 m^2 - s \right ) \nonumber \\
&+ \sqrt f \left ( \tanh^{-1} \sqrt f - i \pi / 2 \right ) \theta \left ( s - 4 m^2 \right ),
\label{JS}
\end{align}
with $\sqrt f = ( 1 - 4 m^2 / s )^{1 / 2}$ and $\theta$ denoting the Heaviside step function. The first term of $J$ in (\ref{JS}) is the analytic continuation of $J$ for the center-of-mass energy below the threshold. To obtain a closed form for $\Pi_{K \overline{K}}$, we set $m = m_K$ and $\epsilon = 1$ in (\ref{RPi}) and for $\Pi_{\pi \pi}$ setting $m = m_\pi$ and $\epsilon = 1 / 2$. Now, all the $T^I_{i j}$ that have been introduced in (\ref{e:T11}) can be expressed in closed form. The parameter $\Lambda$ is fixed as $\Lambda = 1351$ MeV \cite{SPK2011, SPK2025}, so that the real part of $T^0_{1 1}$'s pole correctly reproduces the observed mass of $f_0(980)$. We point out that, having meson--meson scattering amplitudes at our disposal, the respective amplitudes for meson pair production from photon--photon collisions can be evaluated as well; see \cite{JAO1998, SPK2025} for a detailed discussion. These amplitudes are \cite{SPK2025},
\begin{equation}
T_{\gamma \gamma \to m_1 m_2} = \frac{i e^2}{8 \pi^2} \sum_{m_\pm} J_{m_\pm} T_{m_+ m_-; m_1 m_2},
\label{e:FSIhel}
\end{equation}
where $m_\pm$ denotes the masses of the propagating charged meson pairs in the loop, $T_{m_+ m_-; m_1 m_2}$ can be obtained from (\ref{e:T11}), and $J_{m_\pm}$ is given by
\begin{align}
&J_{m_\pm} = \left [ 1 - \frac{4 m^2_\pm}{s} \left ( \sin^{-1} \sqrt{\frac{s}{4 m^2_\pm}} \right )^2 \right ] \nonumber \\
&\times \theta \left ( 4 m^2_\pm - s \right ) + \Biggl [ 1 + \frac{4 m^2_\pm}{s} \nonumber \\
&\times \biggl ( \cosh^{-1} \sqrt{\frac{s}{4 m^2_\pm}} - i \pi / 2 \biggr )^2 \Biggr ] \theta \left ( s - 4 m^2_\pm \right ).
\label{e:JpiK2}
\end{align}

Using the formalism summarized above, combined with QED, we have studied cross-sections for meson pair production from photon--photon collisions in \cite{SPK2025}. However, in \cite{SPK2025}, as in the above summary, we have neglected the possible formation of kaonium, which requires incorporating isospin breaking into the formalism. In the following, we implement the effects of isospin breaking in the amplitudes and recalculate the cross-sections for the two processes $\gamma \gamma \to \pi^0 \pi^0$ and $\gamma \gamma \to \pi^0 \eta$, including the formation of kaonium.

\subsection{Cross-section for $\gamma \gamma \to \pi^0 \pi^0$ including the formation of kaonium}
\label{3A}

The cross-section for $\gamma \gamma \to \pi^0 \pi^0$ is given by
\begin{align}
\sigma \left( \gamma \gamma \to \pi^0 \pi^0 \right) = &\frac{1}{256 \pi^2 s} \frac{p_\pi}{q_i} \nonumber \\
&\times \int_{2 \pi} d \Omega_f \sum_{\lambda \lambda^{'}} \left| \epsilon_\lambda^\mu (1) T_{\mu \nu} \epsilon_{\lambda^{'}}^\nu (2) \right|^2,
\label{siggp}
\end{align}
where the incoming photon and the outgoing pion momenta read $q_i = \sqrt{s} / 2$ and $p_\pi = \sqrt{s} \sqrt{1 - 4 m_\pi^2 / s} / 2$, respectively, and $\epsilon$ denotes the polarization of the photon. After averaging over the photon pair polarization and integrating over half the solid angle of one of the outgoing identical $\pi^0$'s, (\ref{siggp}) becomes
\begin{align}
&\sigma \left( \gamma \gamma \to \pi^0 \pi^0 \right) = \frac{1}{256 \pi s} \sqrt{1 - \frac{4 m_\pi^2}{s}} \left( \frac{\alpha}{\pi} \right)^2 \nonumber \\
&\times \left| J_\pi T_{\pi^+ \pi^-; \pi^0 \pi^0} + J_K T_{K^+ K^-; \pi^0 \pi^0} \right|^2,
\label{crp0}
\end{align}
where $J_\pi$ and $J_K$ can be obtained from (\ref{e:JpiK2}). The charge exchange amplitude for $s$-wave pions in terms of the $\pi \pi$ amplitudes of good isospin $I = 0, 2$ in the isospin basis is $T_{\pi^+ \pi^-; \pi^0 \pi^0} = \langle \pi^0 \pi^0 | T | \pi^+ \pi^- \rangle = ( T_{2 2}^0 - T_{2 2}^2 ) / 3$ and the annihilation amplitude into two neutral pions reads $T_{K^+ K^-; \pi^0 \pi^0} = \langle \pi^0 \pi^0 | T | K^+ K^- \rangle = T_{2 1}^0 / \sqrt{6}$. With these two amplitudes, (\ref{crp0}) takes the form
\begin{align}
&\sigma \left( \gamma \gamma \to \pi^0 \pi^0 \right) = \frac{1}{256 \pi s} \sqrt{1 - \frac{4 m_\pi^2}{s}} \left( \frac{\alpha}{\pi} \right)^2 \nonumber \\
&\times \left| J_\pi \frac{1}{3} \left( T_{2 2}^0 - T_{2 2}^2 \right) + J_K \frac{1}{\sqrt{6}} T_{2 1}^0 \right|^2.
\label{crp0c}
\end{align}

However, note that the above results do not consider isospin violation. Thus, (\ref{crp0c}) cannot be the correct formula for the cross-section if the formation of kaonium is to be considered. The amplitude $T_{\pi^+ \pi^-; \pi^0 \pi^0}$ is not affected by the isospin breaking in the $K^+ K^-$ channel, and thus the effects of isospin breaking due to the $K^0 - K^\pm$ mass difference and $K^+ K^-$ Coulomb interactions should be imported into $T_{K^+ K^-; \pi^0 \pi^0}$. The correct formula for $T_{K^+ K^-; \pi^0 \pi^0}$ should contain the value of the modified strong inverse scattering length $\alpha_p = 1 / a_p$ in (\ref{app}) that determines $k \cot \delta$ from (\ref{cottd}). Note that suppressing the Coulomb field, for example, leaves us with the expected result $k \cot \delta = - \alpha_p$ that only includes the kaon mass difference effects but no kaonium formation. To obtain the correct form of $T_{K^+ K^-; \pi^0 \pi^0}$, we follow the approach proposed in \cite{RHD1959, RHD1960}, which we have already used in Sec.~\ref{be} to import the effects of isospin breaking for the correct calculation of the binding energy of kaonium. The basic physical assumption in those calculations is that isospin is effectively conserved inside the strong-interaction zone; in other words, the isospin-breaking agents are simply too weak with respect to strong interactions to have a significant effect. We can make this remark quantitative by examining the strong-interaction matrix elements in the particle basis involving kaons; they all contain an additive kaon mass squared. Thus, they differ by $(m_{K^0}^2 - m_{K^\pm}^2) / f^2 \approx 2 m_K \Delta / f^2 \approx 0.46$ compared with a typical on-shell value $\sim s / f^2 \approx 4 m_K^2 / f^2 \approx 116$, or about $0.4 \%$ correction in the vicinity of the $K \overline{K}$ threshold, implying that the isospin violation occurs effectively only outside the strong-interaction zone. Taking note of these remarks, we assume the following two points: (i) Since we are interested in the energy region $P_0 \sim 2 m_K$ around the kaonium resonance(s), we ignore symmetry-breaking effects in the pion sector. Such effects would only become relevant for a study of pionium that lies just below the two-pion threshold. (ii) In the kaon sector, the strong-interaction matrix elements are unaffected by symmetry-breaking effects; only the $K^+ K^-$ elastic and charge exchange amplitudes are modified as described by (\ref{esT}) and the last relation in (\ref{pst1}). We implement the aforementioned arguments as follows.

First, note that we have the particle-to-isospin basis relations:
\begin{equation}
\begin{aligned}
&V_{K^+ K^-; \pi^0 \pi^0} = \frac{1}{\sqrt{6}} V_{2 1}^0 = \frac{s}{4 f^2}, \\
&T_{K^+ K^-; K^+ K^-} + T_{K^+ K^-; K^0 \overline{K}^0} = T_{1 1}^0, \\
&T_{K^+ K^-; K^+ K^-} - T_{K^+ K^-; K^0 \overline{K}^0} = T_{1 1}^1,
\end{aligned}
\label{ptib}
\end{equation}
where we have used (\ref{onsv}), $T_{K^+ K^-; K^+ K^-} = \langle K^+ K^- | T | K^+ K^- \rangle = (T_{1 1}^0 + T_{1 1}^1) / 2$, and $T_{K^+ K^-; K^0 \overline{K}^0} = \langle K^0 \overline{K}^0 | T | K^+ K^- \rangle = (T_{1 1}^0 - T_{1 1}^1) / 2$. From (\ref{ptib}) and noting that $T_{2 1}^I = V_{2 1}^I (1 + \Pi_{1 1}^I T_{1 1}^I) / (1 - V_{2 2}^I \Pi_{2 2}^I)$, we obtain
\begin{align}
&T_{K^+ K^-; \pi^0 \pi^0} = \frac{1}{\sqrt{6}} V_{2 1}^0 \nonumber \\
&\times \frac{1 + \Pi_{1 1}^0 \left( T_{K^+ K^-; K^+ K^-} + T_{K^+ K^-; K^0 \overline{K}^0} \right)}{1 - V_{2 2}^0 \Pi_{2 2}^0}.
\label{Tnoy}
\end{align}
This expression for the amplitude $T_{K^+ K^-; \pi^0 \pi^0}$ still does not contain any isospin-breaking effects. We use point (ii) discussed above to import these by approximating the $K^+ K^-$ elastic and charge exchange amplitudes by (\ref{esT}) and (\ref{pst1}) that include both the $K^0 - K^\pm$ mass difference and Coulomb isospin breaking,
\begin{equation}
\begin{aligned}
&T_{K^+ K^-; K^+ K^-} \approx T_{\text{el.}} \approx T_p, \; \; \text{(off kaonium resonances)} \\
&T_{K^+ K^-; K^0 \overline{K}^0} \approx T_{\text{c.e.}} = T_n.
\end{aligned}
\label{ibfe}
\end{equation}
The $K \overline{K}$ polarization loop, $\Pi_{1 1}^0 = \Pi_{K \overline{K}}$, also depends on the mass of the kaon circulating the loop; thus,
\begin{equation}
\Pi_{K^+ K^-} \neq \Pi_{K^0 \overline{K}^0}.
\label{polcr}
\end{equation}
However, except for having thresholds at different branch points $s = 4 m_{K^\pm}^2 < 4 m_{K^0}^2$, their difference in value is $\mathcal{O}(1 \%)$. The modified version of the $K^+ K^- \to \pi^0 \pi^0$ amplitude given by (\ref{Tnoy}) for insertion into (\ref{crp0}) to calculate the $\gamma \gamma \to \pi^0 \pi^0$ cross-section then reads
\begin{align}
T_{K^+ K^-; \pi^0 \pi^0} \approx &\frac{1}{\sqrt{6}} V_{2 1}^0 \nonumber \\
&\times \frac{1 + \Pi_{K^+ K^-} T_\text{el.} + \Pi_{K^0 \overline{K}^0} T_{\text{c.e.}}}{1 - V_{2 2}^0 \Pi_{2 2}^0}.
\label{ftc}
\end{align}

\begin{figure}

\centering

\includegraphics[scale = 0.75]{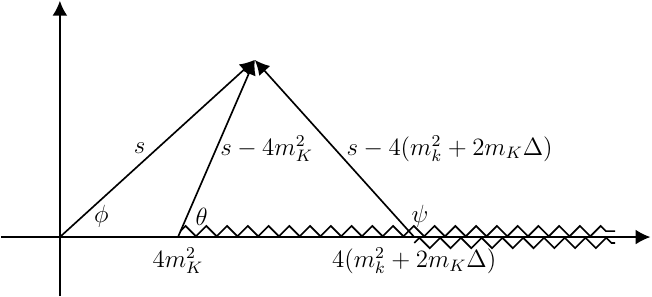}

\caption{Illustration of complex vectors $s$, $s - 4 m_K^2$, and $s - 4 (m_K^2 + 2 m_K \Delta)$, where $-\pi < \phi < \pi$ and $0 < (\theta, \psi) < 2 \pi$. The zigzag lines represent the branch cuts.}

\label{cutpl}
\end{figure}

A couple of remarks are in order: (a) In actual calculations in the vicinity of the kaonium resonances, we approximate $\Pi_{K^+ K^-}$ and $\Pi_{K^0 \overline{K}^0}$ by the common polarization $\Pi_{1 1}^0 = \Pi_{K \overline{K}}$ evaluated at the average kaon mass $m_K$, thus (\ref{ftc}) reduces to:
\begin{equation}
T_{K^+ K^-; \pi^0 \pi^0} \approx \frac{1}{\sqrt{6}} V_{2 1}^0 \frac{1 + \Pi_{K \overline{K}} \left( T_\text{el.} + T_{\text{c.e.}} \right)}{1 - V_{2 2}^0 \Pi_{2 2}^0}.
\label{TKPi}
\end{equation}
(b) To study the influence of the kaonium resonances on the $\gamma \gamma \to \pi^0 \pi^0$ (and later for $\gamma \gamma \to \pi^0 \eta$) cross-section, we need to analytically continue $T_{\text{el.}}$ and $T_{\text{c.e.}}$ to include the unphysical region $0 < s < 4 m_K^2$ of the real $s$ axis where these resonances occur. Both amplitudes are analytic in the upper half complex $s$ plane, cut along the real axis from their respective branch points (thresholds) at $s \approx 4 m_K^2$ and $\approx 4 (m_K^2 + 2 m_K \Delta)$ to $+\infty$. Hence, the analytic extension onto the real axis lying below their respective branch points is given by
\begin{widetext}
\begin{equation}
\begin{aligned}
&T_{\text{el.}} = \frac{8 \pi \sqrt{s}}{k \cot \delta - i k} \approx T_p = 8 \pi \sqrt{s} \frac{-(a_0 + a_1) / 2 - i k_n a_0 a_1}{1 + i (k + k_n) (a_0 + a_1) / 2 - k k_n a_0 a_1} \quad \text{(off kaonium resonances)} \\
&\approx T_{K^+ K^-} = -\frac{1}{2 M_{\text{Ka.}}} \frac{g_{\text{Ka.}}^2}{P_0 - M_{\text{Ka.}} + i \Gamma_{\text{Ka.}} / 2}, \quad \text{(near kaonium resonances, $\sqrt{s} \sim M_{\text{Ka.}}$)} \\
&T_{\text{c.e.}} = T_n = 8 \pi \sqrt{s} \frac{-(a_0 - a_1) / 2}{1 + i (k + k_n) (a_0 + a_1) / 2 - k k_n a_0 a_1}, \\
&T_{\text{e.l.}} + T_{\text{c.e.}} \approx T_p + T_n = 8 \pi \sqrt{s} \frac{-a_0 - i k_n a_0 a_1}{1 + i (k + k_n) (a_0 + a_1) / 2 - k k_n a_0 a_1},
\end{aligned}
\label{anff}
\end{equation}
\end{widetext}
where
\begin{equation}
\begin{aligned}
&k = + \frac{i}{2} \sqrt{4 m_K^2 - s}, \\
&k_n = + \frac{i}{2} \sqrt{4 \left( m_K^2 + 2 m_K \Delta \right) - s},
\end{aligned}
\label{kknan}
\end{equation}
with $0 < s \leq 4 m_K^2$ for $k$ and $0 < s < 4 (m_K^2 + 2 m_K \Delta)$ for $k_n$. To see how (\ref{kknan}) holds, note that $s = |s| e^{i \phi}$ and $s - 4 m_K^2 = |s - 4 m_K^2| e^{i \theta}$, where $-\pi < \phi < \pi$ and $0 < \theta < 2 \pi$ define the first Riemann sheet. Now, $k = (s - 4 m_K^2)^{1 / 2} / 2 = \sqrt{|s - 4 m_K^2|} e^{i \theta / 2} / 2 \to \sqrt{4 m_K^2 - s} e^{i \pi / 2} / 2 = + i \sqrt{4 m_K^2 - s} / 2$, as $\phi \to 0$ and $\theta \to \pi$. Similarly, $s - 4 (m_K^2 + 2 m_K \Delta) = |s - 4 (m_K^2 + 2 m_K \Delta)| e^{i \psi}$, where $0 < \psi < 2 \pi$. As $\phi \to 0$ and $\psi \to \pi$, $k_n = \sqrt{|s - 4 (m_K^2 + 2 m_K \Delta)|} e^{i \psi / 2} / 2 \to \sqrt{4 (m_K^2 + 2 m_K \Delta) - s} e^{i \pi / 2} / 2 = + i \sqrt{4 (m_K^2 + 2 m_K \Delta) - s} / 2$. The cut plane is shown in Fig.~\ref{cutpl}.

It is instructive to study the common complex pole of $T_p$ and $T_n$ given by (\ref{anff}) in the complex $s$ plane. By introducing the complex variable $z = s / 4 m_K^2$, the complex wave numbers (\ref{kknan}) that appear in the denominators of these two amplitudes read
\begin{equation}
\begin{aligned}
k(z) &= m_K \sqrt{\left| z - 1 \right|} e^{i \zeta}, \\
k_n(z) &= m_K \sqrt{\left| z - \left( 1 + 2 \Delta / m_K \right) \right|} e^{i \xi}.
\end{aligned}
\label{ankf}
\end{equation}
The pole of either $T$ is then given by the complex root of the function:
\begin{align}
D(z) = 1 &+ i \left[ k(z) + k_n(z) \right] (a_0 + a_1) / 2 \nonumber \\
&- k(z) k_n(z) a_0 a_1.
\label{rcd}
\end{align}
Now, for the total complex energy of the system, we can write: $M_0 - i \Gamma_0 / 2 \approx 2 m_K + k^2 (z_r) / m_K$, where $D(z_r) = 0$. For the values of $a_0$ and $a_1$ given in \cite{SPK2011}, we can solve for $z_r$, and obtain $M_0 - i \Gamma_0 / 2 \approx (982 - 26 i)$ MeV, which is typical for the resonance $f_0 (980)$.

\begin{figure*}

\centering

\includegraphics[scale = 1.1]{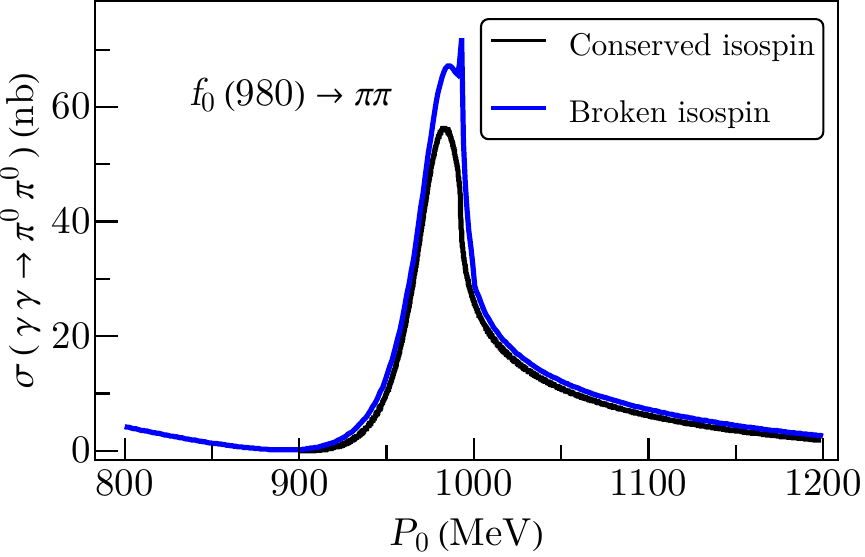}

\caption{Cross-section for $\gamma \gamma \to \pi^0 \pi^0$ as a function of the center-of-mass energy $\sqrt{s} = P_0$. The black curve represents $\sigma ( \gamma \gamma \to \pi^0 \pi^0 )$ with conserved isospin, corresponding to (\ref{crp0c}). The blue curve represents $\sigma ( \gamma \gamma \to \pi^0 \pi^0 )$ with broken isospin, corresponding to (\ref{crp0}) with $T_{K^+ K^-; \pi^0 \pi^0}$ given in (\ref{TKPi}). The isospin breaking is only significant in the close vicinity of the kaonium resonance around 992 MeV. The cutoff parameter is $\Lambda = 1351$ MeV.}

\label{crp00}
\end{figure*}

We now have all the required formulas at our disposal to calculate the cross-section for the process $\gamma \gamma \to \pi^0 \pi^0$. We have shown the result as a function of the center-of-mass energy $P_0$ in Fig.~\ref{crp00}. In this figure, the blue curve, for which the formation of kaonium is taken into account---(\ref{crp0}) is evaluated using $T_{K^+ K^-; \pi^0 \pi^0}$ given in (\ref{TKPi})---shows the same behavior in most of the energy range as the black curve representing (\ref{crp0c}), where the isospin is conserved. The isospin violation manifests itself around 992 MeV as a sharp peak in the close vicinity of the kaonium resonance. In these calculations, we have fixed the cutoff parameter $\Lambda$ at 1351 MeV and have assumed $a_0 = (3.530 - 2.077 i) m_K^{-1}$ and $a_1 = (1.651 - 1.416 i) m_K^{-1}$ in line with our previous work in \cite{SPK2011, SPK2025}.

\subsection{Cross-section for $\gamma \gamma \to \pi^0 \eta$ including the formation of kaonium}
\label{3B}

The expression for the cross-section of the process $\gamma \gamma \to \pi^0 \eta$ differs from that of two-pion production in (\ref{crp0}) in four respects: (i) The angular integral runs over the full solid angle. (ii) The outgoing $\pi^0$ or $\eta$ momentum is $p_{\pi \eta} = \sqrt{s} \sqrt{1 - (m_\pi - m_\eta)^2 / s} \sqrt{1 - (m_\pi + m_\eta)^2 / s} / 2$. (iii) $T_{\pi^+ \pi^-; \pi^0 \eta} = 0$ from $G$-parity conservation; since $\pi^\pm$ carry $I^G (J^P) = 1^- (0^-)$ and $\pi^0$ carries $I^G (J^{P C}) = 1^- (0^{- +})$, while $\eta$ carries $0^+ (0^{- +})$. (iv) $T_{K^+ K^-; \pi^0 \pi^0} \to T_{K^+ K^-; \pi^0 \eta}$. Hence,
\begin{align}
&\sigma \left( \gamma \gamma \to \pi^0 \eta \right) = \frac{1}{128 \pi s} \nonumber \\
&\times \sqrt{1 - \frac{\left( m_\pi - m_\eta \right)^2}{s}} \sqrt{1 - \frac{\left( m_\pi + m_\eta \right)^2}{s}} \left( \frac{\alpha}{\pi} \right)^2 \nonumber \\
&\qquad \qquad \qquad \qquad \quad \times \left| J_K T_{K^+ K^-; \pi^0 \eta} \right|^2.
\label{crpe}
\end{align}
To evaluate $T_{K^+ K^-; \pi^0 \eta}$, if we assume $T_{K^+ K^-; \pi^0 \eta} = \langle \pi^0 \eta | T | K^+ K^- \rangle = - T_{2 1}^1 / \sqrt{2}$, then (\ref{crpe}) becomes
\begin{align}
&\sigma \left( \gamma \gamma \to \pi^0 \eta \right) = \frac{1}{128 \pi s} \nonumber \\
&\times \sqrt{1 - \frac{\left( m_\pi - m_\eta \right)^2}{s}} \sqrt{1 - \frac{\left( m_\pi + m_\eta \right)^2}{s}} \left( \frac{\alpha}{\pi} \right)^2 \nonumber \\
&\qquad \qquad \qquad \qquad \quad \times \left| J_K \frac{1}{\sqrt{2}} T_{2 1}^1 \right|^2.
\label{crpec}
\end{align}

As was the case with the two-pion production cross-section in (\ref{crp0c}), (\ref{crpec}) also does not contain any isospin-breaking effects. To import these, we follow the procedure explained in Sec.~\ref{3A}. First, considering (\ref{ptib}) together with $V_{K^+ K^-; \pi^0 \eta} = - V_{2 1}^1 / \sqrt{2}$, we obtain
\begin{align}
&T_{K^+ K^-; \pi^0 \eta} = -\frac{1}{\sqrt{2}} V_{2 1}^1 \nonumber \\
&\times \frac{1 + \Pi_{1 1}^1 \left( T_{K^+ K^-; K^+ K^-} - T_{K^+ K^-; K^0 \overline{K}^0} \right)}{1 - V_{2 2}^1 \Pi_{2 2}^1}.
\label{tkpeta}
\end{align}
Then, following the arguments presented in Sec.~\ref{3A}, the modified version of the $K^+ K^- \to \pi^0 \eta$ amplitude required to calculate the $\gamma \gamma \to \pi^0 \eta$ cross-section in (\ref{crpe}), when the formation of kaonium is taken into account, can be written as
\begin{equation}
T_{K^+ K^-; \pi^0 \eta} \approx -\frac{1}{\sqrt{2}} V_{2 1}^1 \frac{1 + \Pi_{K \overline{K}} \left( T_\text{el.} - T_{\text{c.e.}} \right)}{1 - V_{2 2}^1 \Pi_{2 2}^1}.
\label{tfke}
\end{equation}
The necessary analytic continuation of $T_{\text{el.}}$ and $T_{\text{c.e.}}$ for studying the influence of the kaonium resonances on the cross-section is the same as before; see (\ref{anff}) and (\ref{kknan}).

\begin{figure*}

\centering

\includegraphics[scale = 1.1]{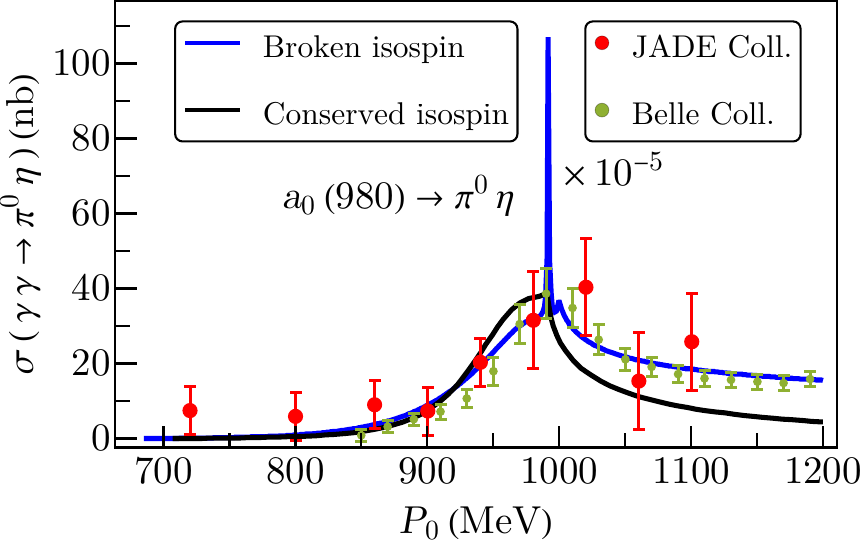}

\caption{Cross-section for $\gamma \gamma \to \pi^0 \eta$ as a function of the center-of-mass energy $\sqrt{s} = P_0$. The black curve represents $\sigma ( \gamma \gamma \to \pi^0 \eta )$ with conserved isospin, corresponding to (\ref{crpec}). The blue curve represents $\sigma ( \gamma \gamma \to \pi^0 \eta )$ with broken isospin, corresponding to (\ref{crpe}) with $T_{K^+ K^-; \pi^0 \eta}$ given in (\ref{tfke}). The kaonium resonance manifests itself as a sharp peak around 992 MeV, where its width is smaller by a factor of $10^{-5}$ than that of the strong-interaction resonance width. The experimental data of the JADE and Belle Collaborations \cite{JADE1990, Belle2009} are shown in red and green, respectively. The cross-section curve, when the formation of kaonium is taken into account, gives a better fit to the experimental data. The cutoff parameter is $\Lambda = 1351$ MeV.}

\label{ggp0et}
\end{figure*}

We are now in a position to evaluate $\sigma ( \gamma \gamma \to \pi^0 \eta )$ as a function of the center-of-mass energy $P_0$. We have shown it in Fig.~\ref{ggp0et} compared to the experimental data from the JADE and Belle Collaborations \cite{JADE1990, Belle2009}. The blue curve corresponds to (\ref{crpe}) with $T_{K^+ K^-; \pi^0 \eta}$ given in (\ref{tfke}) and the black curve represents (\ref{crpec}). Kaonium manifests itself as a sharp resonance around 992 MeV. In contrast to the case of $\gamma \gamma \to \pi^0 \pi^0$, here the difference between the two curves is more significant in the wider energy range. The cross-section curve, when the formation of kaonium is taken into account, gives a better fit to the experimental data.

\subsection{Ratio of the $\gamma \gamma \to \pi^0 \eta$ to $\gamma \gamma \to \pi^0 \pi^0$ cross-sections at the kaonium resonance energy}
\label{3C}

We conclude this section by commenting on the numerical differences between $\gamma \gamma \to \pi^0 \eta$ and $\gamma \gamma \to \pi^0 \pi^0$ at the kaonium resonance energy of $M_{\text{Ka.}} = 991.994$ MeV. The amplitude ratio for the two processes in question at $M_{\text{Ka.}}$ is approximately:
\begin{equation}
\left| \frac{T_{K^+ K^-; \pi^0 \eta}}{T_{K^+ K^-; \pi^0 \pi^0}} \right| \approx 2.6,
\label{rpa}
\end{equation}
where we have used (\ref{TKPi}) and (\ref{tfke}), and have ignored the non-resonant amplitude in the vicinity of the kaonium mass. The value in (\ref{rpa}) can be traced to the difference in behavior of the $\pi \eta$ versus $\pi \pi$ propagators, $\Pi_{2 2}^1$ and $\Pi_{2 2}^0$, as well as the difference in the interactions $V_{2 1}^1$ and $V_{2 1}^0$. We can make this statement quantitative by recalling that $T_{\text{el.}} \gg T_{\text{c.e.}}$ near $s \sim M_{\text{Ka.}}^2$; therefore, all references to the isospin-breaking factors, $T_{\text{el.}} \mp T_{\text{c.e.}}$, drop out of the ratio, and we can write
\begin{align}
\left| \frac{T_{K^+ K^-; \pi^0 \eta}}{T_{K^+ K^-; \pi^0 \pi^0}} \right| &\approx \left| -\sqrt{3} \frac{V_{2 1}^1}{V_{2 1}^0} \frac{\left( 1 - V_{2 2}^1 \Pi_{2 2}^1 \right)^{-1}}{\left( 1 - V_{2 2}^0 \Pi_{2 2}^0 \right)^{-1}} \right| \nonumber \\
&\approx 2.6,
\label{fffn}
\end{align}
confirming explicitly that the difference in value is due to the isospin-conserving strong interactions alone. The resulting cross-section ratio at the kaonium resonance is then
\begin{align}
\frac{\sigma \left( \gamma \gamma \to \pi^0 \eta \right)}{\sigma \left( \gamma \gamma \to \pi^0 \pi^0 \right)} &\approx 2 \frac{p_{\pi \eta}}{p_\pi} \left| \frac{T_{K^+ K^-; \pi^0 \eta}}{T_{K^+ K^-; \pi^0 \pi^0}} \right|^2 \nonumber \\
&\approx 9,
\label{lfinp}
\end{align}
compared to the value of $\approx 8.8$ which can be obtained directly from the ratio of (\ref{crpe}) and (\ref{crp0}) without approximation.

\section{Conclusions}
\label{co}

In this paper, we have studied the binding energies of the 1s and 2s bound states of the hypothetical mesonic atom, $K^+ K^-$ (kaonium), using the $K^+ K^- \to K^+ K^-$ elastic scattering amplitude. The resonance peaks of the amplitude, corresponding to the kaonium binding energies, are in line with previously reported values derived from the pole positions of the Kudryavtsev--Popov equation. Employing a version of chiral perturbation theory (ChPT), developed in \cite{JAO1997, JAO1998}, we have evaluated the cross-sections for processes $\gamma \gamma \to \pi^0 \pi^0$ and $\gamma \gamma \to \pi^0 \eta$, including the formation of kaonium. This requires a modification of meson--meson scattering amplitudes to incorporate the effects of isospin breaking, stemming from the kaon mass difference and Coulomb interactions---the latter being essential for kaonium formation. Our calculations demonstrate that this exotic atom, with a ground-state lifetime of $\sim 10^{-18}$ s, manifests itself as a sharp peak around 992 MeV accompanying $f_0 (980)$ or $a_0 (980)$. The obtained theoretical cross-section curves---particularly for the process $\gamma \gamma \to \pi^0 \eta$---show a better fit to the experimental data from the JADE and Belle Collaborations \cite{JADE1990, Belle2009} when kaonium formation is included in the calculations, and thus hint at the likely existence of this elusive, yet-to-be-observed atom. Building on this improved agreement with the data, we conclude by commenting on the experimental testability of the kaonium signal predicted in this work. The intrinsic width of the ground state of kaonium is extremely small, $\Gamma_{\text{Ka.}} \simeq 0.45$ keV, implying that a direct experimental resolution of the isolated kaonium peak would require a center-of-mass energy resolution $\Delta E_{\text{Res.}} \lesssim 0.45$ keV around 992 MeV. Such a resolution is far beyond the capabilities of current photon–-photon experiments. However, the physical significance of kaonium does not rely on resolving its intrinsic width. Instead, kaonium manifests itself through a localized modification of the energy dependence of the cross-sections near the $K^+ K^-$ threshold, where the background is governed by the much broader $f_0 (980)$ and $a_0 (980)$ resonances. A more realistic regime for near-future facilities is $\Gamma_{\text{Ka.}} \ll \Delta E_{\text{Res.}} \ll \Gamma_{f_0, a_0}$, where the kaonium signal, though broadened by the experimental resolution, appears as a narrow enhancement on a locally flat background. A resolution of $\Delta E_{\text{Res.}} \sim$ 1--2 MeV---which may become accessible with ongoing upgrades to photon--photon experiments---could produce a detectable enhancement confined to 1--2 bins.

\end{document}